\documentclass[conference,letterpaper,10pt]{IEEEtran}

\usepackage[T1]{fontenc}
\usepackage{mathptmx}      
\usepackage{amsmath}
\usepackage{array}
\usepackage{booktabs}
\usepackage[table]{xcolor} 
\usepackage{multicol}      
\usepackage{etoolbox}      
\usepackage{balance}       
\usepackage{graphicx}      

\SetSymbolFont{symbols}{normal}{OMS}{cmsy}{m}{n}
\SetSymbolFont{symbols}{bold}{OMS}{cmsy}{b}{n}

\newcommand{\code}[1]{{\fontencoding{OT1}\fontfamily{cmtt}\selectfont #1}}

\usepackage[
  protrusion=true,
  expansion=true,
  tracking=false,
  kerning=false,
  spacing=false,
  stretch=15, shrink=15,
  final]{microtype}

\definecolor{tabrule}{RGB}{17,58,84}      
\definecolor{tabhead}{RGB}{225,236,245}   
\definecolor{tabalt}{RGB}{245,246,247}    
\definecolor{tabpanelA}{RGB}{221,236,236} 
\definecolor{tabpanelB}{RGB}{251,237,224} 
\arrayrulecolor{tabrule}
\newcommand{\rowhead}{\rowcolor{tabhead}}
\newcommand{\rowalt}{\rowcolor{tabalt}}
\newcommand{\rowpanelA}{\rowcolor{tabpanelA}}
\newcommand{\rowpanelB}{\rowcolor{tabpanelB}}
\newcommand{\tabfont}{\footnotesize}
\newcolumntype{L}[1]{>{\raggedright\arraybackslash}p{#1}}

\newsavebox{\tabbox}

\def\IEEEbibitemsep{0pt}

\makeatletter
\patchcmd{\@BAlancecol}
  {\setbox\@leftcolumn=\vbox to \@tempdima{\unvbox1\vfil}}
  {\setbox\@leftcolumn=\vbox{\unvbox1}%
   \ifdim\ht\@leftcolumn<\@tempdima
     \setbox\@leftcolumn=\vbox to \@tempdima{\unvbox\@leftcolumn\vfil}%
   \fi}
  {}{\PackageError{main}{Could not patch balance.sty's \string\@BAlancecol}%
                        {balance.sty changed; re-check the last-page balance}}
\makeatother

\makeatletter
\patchcmd{\thebibliography}{\section*{\refname}}{}{}{%
  \PackageError{main}{Could not suppress IEEEtran's own References heading}%
                     {thebibliography no longer starts with \string\section*}}
\makeatother

\newenvironment{referencepage}
  {\onecolumn                    
   \addtolength{\topskip}{1pt}%
   \setcounter{unbalance}{2}%
   \begin{multicols}{2}[\section*{\refname}]}
  {\end{multicols}}

\begin{document}

\title{Protocol-Sensitive Evaluation of Log Anomaly Detection:
Component Costs and Target-Access Sensitivity on HDFS and BGL}

\author{%
\IEEEauthorblockN{Hang Xiao}
\IEEEauthorblockA{\textit{Fortinet, Inc.}\\
Sunnyvale, USA\\
xhang@fortinet.com}
\and
\IEEEauthorblockN{Janet Sung}
\IEEEauthorblockA{\textit{Google LLC}\\
Mountain View, USA\\
janetsung@google.com}
\and
\IEEEauthorblockN{Zhaoyi Li}
\IEEEauthorblockA{\textit{Independent Researcher}\\
Mountain View, USA\\
lzy9776@gmail.com}
\and[\hfill\mbox{}\par\vspace{1em}\mbox{}\hfill]
\IEEEauthorblockN{Gangzhen Qian}
\IEEEauthorblockA{\textit{Google LLC}\\
Mountain View, USA\\
irisqian@google.com}
\and
\IEEEauthorblockN{Chuhong Xu}
\IEEEauthorblockA{\textit{Sony Corporate of America}\\
San Jose, USA\\
chuhong.xu@sony.com}
}

\IEEEoverridecommandlockouts
\newsavebox{\arxivnoticebox}
\sbox{\arxivnoticebox}{\parbox[b]{\textwidth}{\footnotesize\raggedright
\textcopyright\ 2026 IEEE. Personal use of this material is permitted.
Permission from IEEE must be obtained for all other uses, in any current or
future media, including reprinting/republishing this material for advertising
or promotional purposes, creating new collective works, for resale or
redistribution to servers or lists, or reuse of any copyrighted component
of this work in other works.}}
\makeatletter
\def\@IEEEpubidpullup{\dimexpr\ht\arxivnoticebox+\dp\arxivnoticebox+12pt\relax}
\makeatother
\IEEEpubid{\usebox{\arxivnoticebox}}
\IEEEaftertitletext{\vspace{-34pt}}
\maketitle

\begin{abstract}
Protocol choices can change the conclusions drawn from log anomaly detection
benchmarks even when detector settings are fixed. We present a joint empirical
study of split construction, representation visibility, and component costs
using six fixed count, sequence, and semantic configurations on Hadoop
Distributed File System (HDFS) and Blue Gene/L (BGL) logs. Random splits place
several configurations near the average-precision ceiling, whereas
group-disjoint HDFS and chronological BGL evaluation produce lower scores and
different observed orderings. At a fixed BGL cutoff, parser choice spans 0.124
in semantic XGBoost mean average precision while preserving its lead over count
XGBoost; the earliest rolling period reverses that ordering. A two-factor
cross-system ablation contrasts source-only representations with offline
transductive access to unlabeled target templates through the representation
corpus and inverse document frequency: HDFS-to-BGL mean average precision moves
from 0.191 with source-only access to 0.325 with union-corpus, target-IDF access,
and the intermediate conditions reveal
direction-dependent interactions in average precision and retrieval at fixed
review budgets. Component-level profiling separates parsing and representation
costs from classifier training, prediction, and storage. Together, these
findings connect detector comparisons to the test population, preprocessing
state, visible information, and measured pipeline stages, and identify the
protocol fields needed alongside a score to support interpretable comparisons
of log anomaly detection accuracy and resource use.
\end{abstract}

\vspace{10pt}

\begin{IEEEkeywords}
log anomaly detection, dependability,
empirical evaluation, evaluation protocol,
target access, computational cost
\end{IEEEkeywords}

\section{Introduction}
\label{sec:intro}

During incident response, an operator usually needs a short list of sessions or
windows to inspect, not another stream of raw messages. A detector therefore has
two practical jobs: rank suspicious activity near the top and keep the review
queue manageable. Benchmark results influence both, so the protocol used to
obtain a score matters.

Early log-mining systems turned console messages into event features; the HDFS
study by Xu and colleagues introduced the dataset used
here~\cite{xu2009detecting}. In our HDFS data, most sessions repeat one of a few
event-count patterns, so a random split places familiar behavior on both sides
of the boundary. He and colleagues organized log anomaly detection as a
pipeline of parsing, feature extraction, and detection~\cite{he2016experience}.
\IEEEpubidadjcol

Such models are evaluated on the same small group of public logs, yet their
scores are hard to compare because the settings differ. DeepLog shifted
attention to event order~\cite{du2017deeplog} and LogAnomaly added semantic
template vectors~\cite{meng2019loganomaly}.

We examine three protocol choices that shape benchmark comparisons: split
construction, the information available during representation building
(target access), and cost accounting. Six configurations with fixed
hyperparameters provide a common basis for examining these choices.
Sedl\'a\v{c}ek and colleagues compared shuffled and sequential evaluation and
preprocessing choices within a unified pipeline~\cite{sedlacek2025anomaly}.
Building on this protocol perspective, our study combines a target-catalog and
inverse document frequency (IDF) visibility ablation, tie-aware review-budget
metrics, and component-level cost accounting. All six configurations
run under one implementation, using a stratified random split, an HDFS holdout
keeping sessions with identical event-count vectors together, and a chronological BGL split with a
parser snapshot at the cutoff. Further runs vary the BGL parser and cutoff, and
two Word2vec corpora cross
\mbox{two IDF sources in both directions.}

The study makes three contributions:
\begin{enumerate}
\item It shows how random, group-disjoint, and chronological evaluation change
      detector comparisons under a common implementation, and characterizes
      parser and temporal-cutoff sensitivity on BGL.
\item It separates parsing, representation construction, classifier training,
      and prediction costs, and reports serialized classifier sizes to make
      the resource requirements of the measured stages explicit.
\item It provides a $2 \times 2$ target-catalog and IDF visibility ablation with
      tie-aware precision and recall at fixed review budgets, stating for each
      cell whether the representation saw unlabeled target artifacts.
\end{enumerate}
Every score below is reported with its split, visible catalog, and measured cost component, so each observation stays tied to \mbox{the protocol that produced it.}

\section{Background and related work}
\label{sec:background}

\subsection{Log parsing}

Log parsers replace identifiers, addresses, counters, and other variable fields
with event templates, fixing the vocabulary seen downstream. Drain organizes
candidate templates in a fixed-depth tree and provides the basis for our compact
BGL parser~\cite{he2017drain}; we use depth 4, similarity threshold 0.4, and
maximum fan-out 100. HDFS retains its released preprocessing while BGL is parsed
from raw messages. Khan and colleagues showed that the parser propagates into
detection accuracy and that parser accuracy alone does not predict downstream
behavior~\cite{khan2024impact}; our BGL sensitivity runs
\mbox{vary this choice for the same reason.}

\subsection{Detection models}

Classical detectors commonly operate on event counts. Sequence models treat
templates as tokens and learn their order, while semantic models retain words
from the template text; the boundaries are loose, and a sequential classifier
may consume semantic event vectors. Ali and colleagues evaluated traditional and deep
techniques in terms of detection accuracy, time performance, and sensitivity to
hyperparameter settings~\cite{ali2025comprehensive}. We use fixed count, sequence, and semantic configurations to compare how each
representation responds to \mbox{split and preprocessing choices.}

\subsection{Evaluation protocols and benchmark critique}

A parallel line of work questions the benchmarks themselves. Landauer and
colleagues surveyed the datasets, preprocessing, models, and evaluation
procedures used across log-anomaly studies and found broad
variation~\cite{landauer2023deep}, then reviewed the common sequence-based data
sets in detail, documenting labeling conventions and structural properties that shape the evaluation tasks these corpora
support~\cite{landauer2024critical}.

Our study fixes the detectors and names the protocol dimensions that differ.
Le and Zhang
showed that training selection, grouping, class distribution, and noise
materially change reported results~\cite{le2022log}, and the study of
Sedl\'a\v{c}ek and colleagues is closest to ours~\cite{sedlacek2025anomaly}.

\subsection{Cross-system anomaly detection}

Cross-system detection adds a vocabulary shift to the usual distribution shift.
LogTransfer trains a model across two systems~\cite{chen2020logtransfer}, and
LogDLR learns a domain-invariant latent representation for unsupervised
cross-system detection~\cite{zhou2025logdlr}. Our experiment examines representation construction: the classifier family,
hyperparameters, and 100-dimensional representation size are fixed while the
Word2vec corpus and IDF source vary. This design measures how target-catalog
\mbox{visibility changes the cross-system feature space.}

\section{Evaluated configurations}
\label{sec:configurations}

The six configurations span count, sequence, and semantic representations
under fixed settings, enabling comparisons across protocol conditions. Each HDFS block session or BGL window is encoded in
three ways. The count representation has one coordinate per template and records
occurrences, dropping event order but keeping the frequency profile; logistic
regression (LR) is the linear baseline, and random forest (RF) and XGBoost the
two tree ensembles. The sequence representation preserves the ordered template
IDs, paired with a single-layer LSTM using a 32-dimensional embedding, 64 hidden
units, and class-balanced loss; its vocabulary produces 28,161 parameters for
HDFS, 115,105 for the BGL random split, and
\mbox{76,033 for the frozen-past BGL run.}

The semantic representation, called Template2Vec here, is shared across systems.
We tokenize the templates and fit a 100-dimensional skip-gram Word2vec
model~\cite{mikolov2013word2vec}; word coordinates retain the same
dimensionality in HDFS and BGL. For a template $t$, we compute:
\begin{equation*}
\mathbf{e}_t = \frac{\sum_{w \in t} \mathrm{IDF}(w) \phi(w)}
                    {\sum_{w \in t} \mathrm{IDF}(w)},
\end{equation*}
when at least one token is in the vocabulary; otherwise the template receives a
zero vector. The session vector is the count-weighted mean of its template
vectors, and LR and XGBoost are trained on that representation.

For the union-corpus cells, Word2vec is fitted on the unlabeled union of
source and target template catalogs. The source-corpus, source-IDF cell uses
source templates alone and is strictly inductive with respect to target-side
access. All runs use a
process-independent SHA-256 word hash, one Word2vec worker, and no external
pretrained model, and five predefined seeds vary
\mbox{fitting and representation initialization.}

\section{Study design}
\label{sec:design}

\subsection{Operational setting and data access}

The study uses an offline benchmark assembled from logs collected by a trusted
monitoring pipeline. Training has access to source logs, templates, and labels. Labels for scored
data remain held out. Inductive runs fit representation models using
training-side artifacts; offline transductive runs additionally use unlabeled
artifacts from the data to be scored. Random and grouped semantic runs share a
corpus over both unlabeled catalogs; in the cross-system ablation, the
source-corpus, source-IDF cell is inductive, and cells using a union corpus or
target IDF are offline transductive. The resulting scores rank benchmark
sessions or windows for review. Section~\ref{sec:discussion} describes the
security assumptions and the operational meaning \mbox{of the released labels.}

\begin{table}[t]
\caption{Dataset characteristics. HDFS and BGL differ in scale, vocabulary,
prevalence, repetition, and session structure.}
\label{tab:datasets}
\centering
\tabfont
\begin{tabular}{lrr}
\toprule
\rowhead Characteristic & HDFS & BGL \\
\midrule
Unit                   & block session  & 100-line window \\
\rowalt Sessions/windows       & 575,061        & 47,479 \\
Anomalies              & 16,838 (2.93\%) & 4,802 (10.11\%) \\
\rowalt Event vocabulary       & 29             & 2,746 \\
Mean length            & 19.4           & 100 \\
\rowalt Distinct count vectors & 589            & 6,280 \\
Repeated-vector share  & 99.96\%        & 88.86\% \\
\bottomrule
\end{tabular}
\end{table}

\subsection{Datasets}

Table~\ref{tab:datasets} summarizes the two datasets. HDFS messages are grouped
by block identifier and inherit the released block-session label. BGL contains
line-level alerts from a Blue Gene/L
supercomputer~\cite{oliner2007supercomputers}. We keep the chronological order,
form non-overlapping 100-line windows, and label a window anomalous when it
contains at least one alert.

HDFS and BGL are historical benchmarks with documented recurrence patterns,
event vocabularies, and labeling units~\cite{landauer2024critical}. Their
contrasting structures provide two settings for examining protocol-dependent
behavior. Extending the same accounting to further labeled Loghub corpora, such
as Thunderbird~\cite{zhu2023loghub}, would broaden \mbox{the empirical coverage.}

Repetition is pronounced in both. HDFS contains 589 count vectors across 575,061
sessions; 99.96\% of sessions share a count vector with another, and 98.09\%
share a complete template-ID sequence. BGL has 6,280 count vectors, with
corresponding shares of 88.86\% and 72.96\%. Under the seed-0 random split,
99.95\% of HDFS test sessions and 88.39\% of BGL test windows match a training
count vector. The group and chronological protocols reduce that recurrence.

\subsection{Research questions and protocols}

RQ1: How do the observed results and component costs change across split
protocols? We begin with a stratified 70/30 random split for each dataset.
HDFS is also evaluated with repeated group-disjoint holdouts: sessions sharing a
count vector form one group, assigned intact to fitting, validation, or test.
BGL supplies timestamps, so its second protocol trains on the first 70\% of
windows and tests on the remaining 30\%; HDFS uses group-disjoint evaluation because its released sessions lack a
reliable global order. The protocols jointly vary the test population and
training--test recurrence; BGL chronological evaluation also changes
\mbox{the calendar interval.}

\begin{table}[t]
\caption{Information available before final test scoring, and the
resulting~access~regime.}
\label{tab:access}
\centering
\savebox{\tabbox}{%
\tabfont
\setlength{\tabcolsep}{3pt}
\begin{tabular}{@{}L{76pt}lll@{}}
\toprule
\rowhead Condition & Labels visible & Template catalog & Access regime \\
\midrule
Within random/group               & own training    & full catalog    & transductive\textsuperscript{*} \\
\rowalt BGL chronological                 & past training   & cutoff snapshot & inductive \\
Cross: source corpus, source IDF  & source training & source only     & inductive \\
\rowalt Cross: union corpus or target IDF & source training & source + target & transductive \\
\bottomrule
\end{tabular}}
\usebox{\tabbox}\par
\vspace{2pt}
\parbox{\wd\tabbox}{\raggedright\tabfont
\textsuperscript{*}\,Scored labels are hidden; transductive access uses unlabeled
scored-data artifacts (Section~\ref{sec:design}).\par}
\end{table}

\begin{table}[t]
\caption{Fixed hyperparameters used across protocols and seeds.}
\label{tab:hyper}
\centering
\tabfont
\setlength{\tabcolsep}{2pt}
\begin{tabular}{@{}lL{178pt}@{}}
\toprule
\rowhead Model & Configuration \\
\midrule
LR       & inverse regularization; balanced weights; 1,000 iterations; standardized input \\
\rowalt RF       & 200 trees; balanced-subsample class weights \\
XGBoost  & 300 trees; depth 6; rate 0.1; histogram method; class-ratio positive weight \\
\rowalt Word2vec & skip-gram; 100 dimensions; window 5; minimum count 1; 30 epochs \\
LSTM     & 32-d embedding; 64 hidden; Adam $10^{-3}$; batch 256; at most five epochs \\
\bottomrule
\end{tabular}
\end{table}

The chronological BGL run saves the parser state at the cutoff, where 1,524
templates are known and the Word2vec and IDF models are fitted. The parser then
advances window by window: count and sequence inputs map new templates to
\textsc{unk}, while semantic inputs use words from the frozen vocabulary and
return zero when none is known. Of the 2,746 templates eventually produced,
1,222 first appear in the test period, and 55.5\% of anomalous test windows
contain at least one. This follows the concern raised by LogRobust: logging code
and observed event patterns change over time~\cite{zhang2019robust}.

RQ2: How stable is the BGL chronological result across parsers and
cutoffs? At the 70/30 cutoff, the compact parser is run with similarity
thresholds 0.3, 0.4, and 0.5, and Drain3 v0.9.11 supplies a fourth configuration
at depth 4 and threshold 0.4; fan-out, masking, and windowing remain fixed. We
then use the default parser for rolling-origin tests at 50\%, 60\%, 70\%, and
80\%, scoring the next 10\% of windows each time. The fixed-cutoff panel varies parser configuration while retaining the other
settings. Rolling-origin evaluation examines temporal scenarios in which
training-prefix size, calendar period, event mix, vocabulary state, and test
interval \mbox{change together.}

RQ3: How does unlabeled target-catalog visibility change cross-system
feature coverage? Two Word2vec corpora (source or union) are crossed with two
IDF sources (source or target); the union corpus and target IDF use target
templates without labels, and Table~\ref{tab:transfer} reports XGBoost in both
directions. Domain-adaptation methods such as LogTAD learn an explicit alignment
across systems~\cite{han2021logtad}; here the four cells keep the classifier family, hyperparameters, labels, and
source training set fixed while varying catalog and IDF visibility. Each cell
constructs its own features and fits a separate classifier. Table~\ref{tab:access} records what is
available and labels each condition inductive or offline transductive; a
union-corpus token absent from source IDF receives
\mbox{the logged fallback weight 1.0.}

\subsection{Training, selection, and repetition}

Each training portion is divided 80/20 into fitting and validation subsets. HDFS
groups remain disjoint, BGL validation follows training in time, and validation
$F_1$ selects the decision threshold. RF follows the random-forest construction
of Breiman~\cite{breiman2001random}; we use the scikit-learn
implementation~\cite{pedregosa2011scikit} with 200 trees and its
\code{balanced\_subsample} class weighting.

Hyperparameters are unchanged across protocols (Table~\ref{tab:hyper}).
XGBoost uses 300 depth-6 trees,
learning rate 0.1, the histogram method, and a class-ratio positive weight,
following the scalable boosted-tree system of Chen and
Guestrin~\cite{chen2016xgboost}. The LSTM trains for at most five epochs and
keeps the epoch with the highest validation AP. HDFS sequences are truncated at 50
events, of which only 0.01\% are longer, and \mbox{BGL keeps all 100 per window.}

Five predefined seeds control the applicable split, group assignment, and
stochastic fitting; the chronological period remains fixed across seeds. HDFS
groups are assigned by a seeded stratified procedure targeting 56/14/30 percent
for fit, validation, and test. Unequal group sizes produce realized test shares
of 13\%--31\% and prevalence of 2.51\%--6.99\%, and groups containing both
classes receive an any-anomaly label during stratified assignment within each
seed.

In Table~\ref{tab:hdfs}, each HDFS group assignment therefore defines a test set
with its own size and anomaly prevalence; the reported means and standard
deviations summarize variation across group assignments \mbox{and model fitting.}

\subsection{Metrics and statistical reporting}

We report anomaly-class precision, recall, and $F_1$ at the validation-selected
threshold. Average precision (AP) is computed with the
\code{average\_precision\_score} function of
\code{sklearn.metrics}~\cite{pedregosa2011scikit}, as a weighted step sum of precision over recall increments. Test prevalence is the random-ranking reference.
Table~\ref{tab:transfer} reports tie-aware precision and recall at 1\% and 5\%
review budgets; a boundary cutting through a score tie receives the expected
fractional allocation under uniform selection within that tied group, and
budgets
\mbox{count benchmark sessions or windows.}

Tables report archived means and standard deviations across five seeds. HDFS
group-holdout dispersion combines variation in group assignments, test
composition, and model fitting; chronological BGL dispersion reflects fitting
within a fixed test period. The archive does not specify whether the
standard-deviation denominator is $n$ or $n-1$. The analysis reports mean
differences and across-run dispersion; inferential tests and confidence
intervals are outside \mbox{the reported analysis.}

\subsection{Cost accounting}

Timing is recorded separately for parsing, representation fitting, feature
transformation, classifier training, and classifier-only inference, and each boundary is reported on its own. Classifier-only throughput is
the median of three passes over the test split, fit time is summarized across
the five seeded runs, and serialized classifier bytes define artifact size:
pickle for the classical configurations, a saved state dictionary for the LSTM.
Word2vec fitting and semantic transformation are timed in the same runs; BGL
parsing is timed once from raw logs, and HDFS begins from released
preprocessing.

Classifier-only throughput measures prediction on prepared features; parsing,
feature construction, model loading, I/O, and analyst delay are outside that
metric, and parsing and representation costs are reported separately. Archived timings were
produced on two threads of an Intel Xeon 2.10 GHz virtual CPU with two cores,
8 GB RAM, Linux, and OpenBLAS. The resource measurements comprise component runtimes and serialized classifier
sizes; peak memory and complete deployment-bundle size were not recorded in the
archive. The Apple runs behind the parser and rolling-origin panels contribute
accuracy results; their detailed timing \mbox{configuration is unavailable.}

\section{Results}
\label{sec:results}

\begin{table}[t]
\caption{HDFS results across five seeds (mean $\pm$ SD); random/group access
is transductive.}
\label{tab:hdfs}
\centering
\tabfont
\setlength{\tabcolsep}{4pt}
\begin{tabular}{lrrrr}
\toprule
\rowhead & \multicolumn{2}{c}{Random split} & \multicolumn{2}{c}{Group-disjoint} \\
\cmidrule(lr){2-3}\cmidrule(lr){4-5}
\rowhead Configuration & \multicolumn{1}{c}{AP} & \multicolumn{1}{c}{$F_1$}
         & \multicolumn{1}{c}{AP} & \multicolumn{1}{c}{$F_1$} \\
\midrule
Count/LR      &  .991$\pm$.004 & .989$\pm$.006 & .867$\pm$.103 & .934$\pm$.046 \\
\rowalt Count/RF      & 1.000$\pm$.000 & .998$\pm$.000 & .743$\pm$.227 & .808$\pm$.259 \\
Count/XGB     &  .999$\pm$.001 & .998$\pm$.000 & .692$\pm$.175 & .582$\pm$.337 \\
\rowalt Sequence/LSTM &  .999$\pm$.001 & .993$\pm$.002 & .627$\pm$.210 & .502$\pm$.324 \\
Semantic/LR   &  .982$\pm$.003 & .981$\pm$.003 & .725$\pm$.174 & .792$\pm$.275 \\
\rowalt Semantic/XGB  & 1.000$\pm$.000 & .998$\pm$.000 & .624$\pm$.235 & .318$\pm$.241 \\
\bottomrule
\end{tabular}
\end{table}

\begin{table}[t]
\caption{BGL results across five seeds (mean $\pm$ SD); random access is
transductive, chronological access is inductive.}
\label{tab:bgl}
\centering
\tabfont
\setlength{\tabcolsep}{4pt}
\begin{tabular}{lrrrr}
\toprule
\rowhead & \multicolumn{2}{c}{Random split} & \multicolumn{2}{c}{Chronological} \\
\cmidrule(lr){2-3}\cmidrule(lr){4-5}
\rowhead Configuration & \multicolumn{1}{c}{AP} & \multicolumn{1}{c}{$F_1$}
         & \multicolumn{1}{c}{AP} & \multicolumn{1}{c}{$F_1$} \\
\midrule
Count/LR      & .987$\pm$.003 & .976$\pm$.004 & .277$\pm$.000 & .423$\pm$.000 \\
\rowalt Count/RF      & .998$\pm$.001 & .984$\pm$.002 & .388$\pm$.005 & .400$\pm$.012 \\
Count/XGB     & .998$\pm$.001 & .989$\pm$.004 & .448$\pm$.000 & .396$\pm$.000 \\
\rowalt Sequence/LSTM & .946$\pm$.012 & .916$\pm$.017 & .236$\pm$.151 & .282$\pm$.091 \\
Semantic/LR   & .916$\pm$.006 & .870$\pm$.007 & .205$\pm$.095 & .267$\pm$.043 \\
\rowalt Semantic/XGB  & .992$\pm$.000 & .955$\pm$.006 & .676$\pm$.086 & .492$\pm$.165 \\
\bottomrule
\end{tabular}
\end{table}

\begin{figure}[t]
\centering
\includegraphics{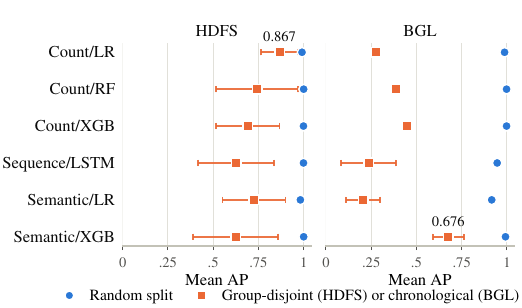}
\caption{Mean AP ($\pm$ standard deviation over five seeds) of the six
configurations under random splitting and each dataset's alternative protocol,
from Tables~\ref{tab:hdfs} and \ref{tab:bgl}. Group-disjoint HDFS and
chronological BGL yield lower mean scores and different leading configurations.}
\label{fig:protocols}
\end{figure}

\subsection{RQ1: Split construction changes the observed ordering}

Tables~\ref{tab:hdfs} and~\ref{tab:bgl} address RQ1; Fig.~\ref{fig:protocols}
plots their mean AP. Random splitting produces near-ceiling AP clusters. On HDFS four
configurations report mean AP within 0.001 of each other, with count/LR at 0.991
and semantic/LR at 0.982 below them; on BGL three configurations lie within
0.006, with count/LR at 0.987, the LSTM at 0.946, and semantic/LR at 0.916 below
them. Count/XGB records AP $0.999 \pm 0.001$ on HDFS and $0.998 \pm 0.001$ on
BGL.
Nearly every HDFS test session and most BGL test windows already match a
training count vector, so the table largely
\mbox{measures interpolation over recurrence.}

Separating the HDFS count-vector groups changes the picture: mean AP falls for
all six configurations and varies far more across assignments, and count/LR records
the highest mean AP at $0.867 \pm 0.103$, followed by RF at $0.743 \pm 0.227$
and semantic/LR at $0.725 \pm 0.174$, with the LSTM and semantic/XGB averaging
0.627 and 0.624. The standard deviations summarize variation across the five group assignments
\mbox{and model fitting.}

The protocol change alters which configuration has the highest reported mean AP. RF and
semantic/XGB have the same reported mean AP in the HDFS random table, whereas their mean AP
differs by 0.119 after grouping, and count/LR moves from below the four
near-ceiling configurations to the highest grouped mean. Its mean $F_1$ falls
from 0.989 to 0.934 while its mean AP falls from 0.991 to 0.867: the
validation-selected operating point and the full ranking respond differently to
the protocol change.

The chronological BGL holdout produces a third ordering, with the period fixed
so seeds vary fitting only. Semantic/XGB records the highest mean AP at
$0.676 \pm 0.086$, count/XGB reaches 0.448, and the LSTM reaches
$0.236 \pm 0.151$. The semantic representation is fitted on the 1,524 templates
available at the cutoff; of the 1,222 appearing later, 418 contain no word from
the frozen vocabulary when first observed and receive a zero vector. Refitting
the representation on the full catalog---an offline transductive variant that
sees post-cutoff templates---changes semantic/XGB AP from $0.676 \pm 0.086$ to
$0.796 \pm 0.053$. Full-catalog access changes zero-vector coverage together with the rest of the representation, so this contrast \mbox{measures their joint effect.}

AP and $F_1$ identify the same highest-scoring BGL configuration; semantic/XGB
has an $F_1$ standard deviation of 0.165. Count/XGB produces identical refits
within this fixed period. The compact LSTM records greater run-to-run variation
and lower AP than the count models. LogBERT provides a transformer-based,
self-supervised sequence \mbox{alternative~\cite{guo2021logbert}.}

\begin{table}[t]
\caption{BGL chronological sensitivity over five seeds.
Parser: 70/30 split; rolling: next 10\% after each cutoff.}
\label{tab:bglsens}
\centering
\tabfont
\setlength{\tabcolsep}{3pt}
\begin{tabular}{@{}llrrrr@{}}
\toprule
\rowhead      & \multicolumn{2}{r}{Test anomaly}
      & \multicolumn{1}{c}{Count/XGB} & \multicolumn{1}{c}{LSTM}
      & \multicolumn{1}{c}{Semantic/XGB} \\
\rowhead Panel & Setting & \multicolumn{1}{c}{rate} & \multicolumn{1}{c}{AP}
      & \multicolumn{1}{c}{AP} & \multicolumn{1}{c}{AP} \\
\midrule
\rowpanelA Parser & Compact $s = .3$ & 10.1\% & .460$\pm$.000 & \multicolumn{1}{c}{--} & .741$\pm$.054 \\
\rowpanelA Parser & Compact $s = .4$ & 10.1\% & .448$\pm$.000 & \multicolumn{1}{c}{--} & .676$\pm$.086 \\
\rowpanelA Parser & Compact $s = .5$ & 10.1\% & .448$\pm$.000 & \multicolumn{1}{c}{--} & .631$\pm$.109 \\
\rowpanelA Parser & Drain3 $s = .4$  & 10.1\% & .432$\pm$.000 & \multicolumn{1}{c}{--} & .617$\pm$.177 \\
\midrule
\rowpanelB Rolling & 50$\rightarrow$60\% &  1.3\% & .902$\pm$.000 & .475$\pm$.147 & .838$\pm$.047 \\
\rowpanelB Rolling & 60$\rightarrow$70\% & 10.8\% & .376$\pm$.000 & .201$\pm$.048 & .663$\pm$.077 \\
\rowpanelB Rolling & 70$\rightarrow$80\% & 13.2\% & .588$\pm$.000 & .373$\pm$.248 & .760$\pm$.103 \\
\rowpanelB Rolling & 80$\rightarrow$90\% &  5.9\% & .332$\pm$.000 & .273$\pm$.130 & .776$\pm$.079 \\
\bottomrule
\end{tabular}
\end{table}

\subsection{RQ2: Parser and temporal sensitivity}

Table~\ref{tab:bglsens} reports the RQ2 experiments. At the fixed 70/30 split
semantic/XGB records a higher mean AP than count/XGB in all four parser
configurations; its mean AP spans 0.617--0.741 and the Drain3 run has a standard
deviation of 0.177. With the cutoff and remaining settings fixed, parser choice spans 0.124 in
semantic/XGB mean AP while preserving the ordering \mbox{of the two configurations.}

The rolling origins vary more. Semantic/XGB records the higher mean AP in the
periods beginning at 60\%, 70\%, and 80\%, while count/XGB is higher at 50\%,
reaching 0.902 AP against 0.838. Test prevalence ranges from 1.3\% to 13.2\%,
and every later training prefix contains the earlier one. Prefix length, calendar period, anomaly rate, and carried-over vocabulary state move together across these rows, so row-to-row differences reflect \mbox{their joint variation.} LogSD also evaluates BGL chronologically, though
its self-supervised model and split differ from ours~\cite{xie2024logsd}.

The two panels provide complementary views: the fixed-cutoff panel isolates
parser sensitivity, while the rolling-origin panel characterizes performance as
training history, calendar period, prevalence, and vocabulary state evolve
together.
SemParser showed that richer parsing semantics can change downstream
analytics~\cite{huo2023semparser} and Khan and colleagues report the same
sensitivity for deep detectors~\cite{khan2024impact}. Even syntax-oriented
thresholds move BGL AP comparably here.

All parser runs use the same raw BGL
release in recorded order, with provenance through Loghub~\cite{zhu2023loghub},
and the 50--60\% rolling period is the single interval in which the order
changes.

\begin{table}[t]
\caption{Component-level cost over five seeds. Throughput covers
classifier-only inference; size covers the serialized classifier.}
\label{tab:cost}
\centering
\tabfont
\setlength{\tabcolsep}{3pt}
\begin{tabular}{@{}llrrr@{}}
\toprule
\rowhead System & Configuration & Fit (s) & Infer. (k/s) & Size (MB) \\
\midrule
\rowpanelA HDFS & Count/LR      &   2.6$\pm$1.1 & 6771$\pm$2151 &  .002 \\
\rowpanelA HDFS & Count/RF      &  17.8$\pm$2.1 &  312$\pm$91   & 2.234 \\
\rowpanelA HDFS & Count/XGB     &   3.8$\pm$1.8 &  660$\pm$94   &  .423 \\
\rowpanelA HDFS & Sequence/LSTM & 309.3$\pm$13.2 &   37$\pm$4   &  .116 \\
\rowpanelA HDFS & Semantic/LR   &  31.4$\pm$21.3 & 1257$\pm$113 &  .004 \\
\rowpanelA HDFS & Semantic/XGB  &   9.4$\pm$.8  &  696$\pm$189  &  .416 \\
\midrule
\rowpanelB BGL & Count/LR      & 16.5$\pm$4.4  &   70$\pm$31   &  .069 \\
\rowpanelB BGL & Count/RF      & 38.9$\pm$18.6 &   29$\pm$3    & 10.869 \\
\rowpanelB BGL & Count/XGB     & 10.2$\pm$3.2  &  222$\pm$74   &  .372 \\
\rowpanelB BGL & Sequence/LSTM & 37.8$\pm$3.9  &   15$\pm$2    &  .386 \\
\rowpanelB BGL & Semantic/LR   &  6.1$\pm$1.7  & 1812$\pm$754  &  .004 \\
\rowpanelB BGL & Semantic/XGB  &  1.6$\pm$.3   &  505$\pm$28   &  .684 \\
\midrule
\multicolumn{2}{l}{\hspace{0.6em}Shared Word2vec fit}          & .4$\pm$.0  & -- & -- \\
\rowalt \multicolumn{2}{l}{\hspace{0.6em}Semantic transform: HDFS/BGL} & 1.0/1.4    & -- & -- \\
\multicolumn{2}{l}{\hspace{0.6em}Frozen BGL semantic fit}      & 3.0$\pm$.5 & -- & -- \\
\rowalt \multicolumn{2}{l}{\hspace{0.6em}BGL parsing, one run}         & 54         & -- & -- \\
\bottomrule
\end{tabular}
\end{table}

\subsection{Where time and storage are spent}

Table~\ref{tab:cost} separates classifier cost from representation work. On
HDFS, count/LR is the smallest serialized classifier and processes 6.77 million
sessions per second on prepared features on the archived hardware; count/XGB occupies 0.423 MB, and the LSTM state is smaller at 0.116 MB
although its fit takes 309.3 seconds against 3.8 for XGBoost. Semantic classifiers achieve the highest classifier-only throughput on BGL;
Word2vec fitting and semantic transformation are
\mbox{reported as separate pipeline stages.}

Storage and runtime emphasize different components: semantic LR uses a
four-kilobyte classifier together with embedding and IDF resources, the compact
LSTM has substantial recurrent-computation cost, and raw BGL parsing adds 54
seconds. Yu and colleagues similarly found lightweight classical methods
competitive when preprocessing and training costs are
included~\cite{yu2024deep}. Component-level accounting makes
\mbox{these resource trade-offs explicit.}

\begin{table}[t]
\caption{Cross-system XGBoost target-catalog visibility over five seeds. Union
corpus and target IDF are offline transductive; budgets use
tie-aware~allocation.}
\label{tab:transfer}
\centering
\tabfont
\setlength{\tabcolsep}{2pt}
\begin{tabular}{@{}llrrrrr@{}}
\toprule
\rowhead Word2vec & IDF & \multicolumn{1}{c}{AP} & \multicolumn{1}{c}{P@1\%}
         & \multicolumn{1}{c}{R@1\%} & \multicolumn{1}{c}{P@5\%}
         & \multicolumn{1}{c}{R@5\%} \\
\midrule
\rowpanelA \multicolumn{7}{c}{HDFS$\rightarrow$BGL} \\
\rowpanelA Source & Source & .191$\pm$.107 & .242$\pm$.160 & .024$\pm$.016 & .220$\pm$.174 & .109$\pm$.086 \\
\rowpanelA Source & Target & .187$\pm$.084 & .259$\pm$.163 & .026$\pm$.016 & .253$\pm$.163 & .125$\pm$.081 \\
\rowpanelA Union  & Source & .297$\pm$.136 & .443$\pm$.266 & .044$\pm$.026 & .293$\pm$.021 & .145$\pm$.010 \\
\rowpanelA Union  & Target & .325$\pm$.159 & .446$\pm$.312 & .044$\pm$.031 & .318$\pm$.122 & .157$\pm$.060 \\
\midrule
\rowpanelB \multicolumn{7}{c}{BGL$\rightarrow$HDFS} \\
\rowpanelB Source & Source & .089$\pm$.066 & .151$\pm$.228 & .051$\pm$.078 & .075$\pm$.057 & .128$\pm$.097 \\
\rowpanelB Source & Target & .132$\pm$.100 & .248$\pm$.297 & .085$\pm$.102 & .095$\pm$.067 & .162$\pm$.114 \\
\rowpanelB Union  & Source & .039$\pm$.017 & .032$\pm$.018 & .011$\pm$.006 & .025$\pm$.007 & .043$\pm$.011 \\
\rowpanelB Union  & Target & .142$\pm$.137 & .201$\pm$.233 & .068$\pm$.080 & .129$\pm$.121 & .221$\pm$.207 \\
\bottomrule
\end{tabular}
\end{table}

\subsection{RQ3: Target-catalog visibility and feature coverage}

Count vectors and event-ID sequences have system-specific dimensions, so
cross-system use needs a template mapping, retraining, or an adaptation model.
Template2Vec has a common dimension and applies directly, exposing the choices
in Table~\ref{tab:transfer}. That table examines cross-system feature coverage and detection scores under
different template-corpus and IDF visibility conditions \mbox{in both directions.}

Between the strictly inductive source-only cell and the fully transductive
union-corpus, target-IDF cell, XGBoost mean AP changes from 0.191
to 0.325 for HDFS-to-BGL and from 0.089 to 0.142 for BGL-to-HDFS. The intermediate cells show direction-dependent interactions between corpus
coverage and IDF weighting: supplying only target IDF or only the
union corpus is associated with a higher mean AP in one direction and a lower
one in the other, and the union-corpus, source-IDF cell for BGL-to-HDFS falls to
0.039, \mbox{below its own source-only value.}

At the union-corpus, target-IDF cell, a 1\% review budget retrieves 4.4\% of BGL
anomalies and 6.8\% of HDFS anomalies on average; a 5\% budget corresponds to
recall of 15.7\% and 22.1\%, with precision of 31.8\% and 12.9\%. These results describe the review queues for each target population under its
stated access condition; Table~\ref{tab:transfer}
\mbox{reports the corresponding dispersion.}

\section{Discussion and threats to validity}
\label{sec:discussion}

\subsection{Reading protocol-sensitive results and their dispersion}

The ordering of six fixed configurations changes with the
protocol, consistent with~\cite{sedlacek2025anomaly}. HDFS group holdout evaluates
unseen count profiles, whereas chronological BGL evaluates a later trace segment
with a frozen representation. These distinct test conditions matter when
comparing benchmark scores~\cite{landauer2023deep}. Changes from random
splitting are therefore interpreted within each dataset's
\mbox{corresponding protocol.}

HDFS dispersion combines variation in test composition and model fitting,
while chronological BGL dispersion reflects refitting within one fixed period
(Section~\ref{sec:design}).

We report differences in mean scores together with their across-run dispersion
and protocol conditions. Four reporting fields make these comparisons
interpretable: the share of test patterns already seen in training for random splits;
the parser snapshot and at least one neighboring period for chronological
splits; the representation corpus and IDF source for cross-system semantic
results; and \mbox{budget recall with run-to-run dispersion.}

\subsection{Preprocessing, parser state, and label semantics}

The parser panel measures downstream detection sensitivity, with a 0.124 span
in semantic/XGB mean AP and an unchanged leading configuration. Template-level
ground-truth annotations are unavailable \mbox{for this BGL evaluation.}

HDFS labels variable-length block sessions, whereas BGL labels fixed 100-line
windows anomalous after one alert. Review-budget precision is therefore measured per block session and per
window. New count/sequence
templates map to \textsc{unk}, while Template2Vec uses known frozen-vocabulary
words or zero. The 418 later templates lacking usable words identify the coverage
boundary of the frozen semantic representation.

\subsection{Temporal estimates as compound scenarios}

The primary BGL experiment evaluates refitting at a 70/30 cutoff, and four
rolling origins extend the analysis across nested training prefixes and
subsequent test periods. The resulting variation is consistent with the
sensitivity to training selection, class distribution, and noise reported by Le
and \mbox{Zhang~\cite{le2022log}.}

The earliest-period reversal shows why evaluation across periods complements
repeated fitting within one period. A factorial extension incorporating methods
such as LogETA~\cite{gong2024logeta} could separate training-prefix and
\mbox{calendar-period effects.}

\subsection{Cross-system results as a coverage diagnostic}

The two directions expose the interaction between corpus structure and
representation access. In the HDFS-to-BGL source-only condition, 91.7\% of
target templates and 67.3\% of target sessions map to zero. Union-corpus
training adds coordinates for those words, and target IDF changes their
weighting. The intermediate conditions show direction-dependent
\mbox{interactions between these choices.}

Related work sits at other points on this axis: MetaLog adapts with a small
labeled target set~\cite{zhang2024metalog}, and LogDLR learns domain-invariant
representations through reconstruction and domain-adversarial
training~\cite{zhou2025logdlr}.
Template2Vec keeps the classifier family, hyperparameters, source labels, and
representation dimension fixed while varying catalog visibility and IDF source;
each condition generates its own feature vectors and \mbox{fits a separate classifier.}

LogGT uses a heterogeneous graph before domain-adapted transfer~\cite{wang2024loggt}.
HDFS contributes 29 templates from regular block sessions, whereas
BGL contributes 2,746 templates from a system-wide stream, providing contrasting
coverage conditions. Additional systems would extend this analysis across
\mbox{other logging styles and failure modes.}

\balance

\subsection{Model scope and cost boundaries}

The six fixed configurations cover count, sequence, and semantic
representations, enabling the same protocol changes to be examined across
distinct feature families. Recent comparative studies provide complementary coverage of
detector families~\cite{ali2025comprehensive} and benchmark
pipelines~\cite{sedlacek2025anomaly}.

The LSTM uses one layer with 64 hidden units and at most five epochs.
Evaluating RT-Log~\cite{jia2023robust}
and LogLLM~\cite{guan2024logllm} under matched split indices, access conditions,
and review budgets \mbox{would extend the model coverage.}

Parser and rolling-origin sensitivity results were computed in a separate Apple
environment; cost measurements in Table~\ref{tab:cost} use the documented
archived \mbox{Xeon environment.}

\subsection{Security scope and threat model}

The evaluation measures failure- and alert-oriented anomaly screening using
released HDFS failure labels and BGL alert flags. CAShift provides a
complementary setting with
\mbox{attack scenarios under normality shift}~\cite{yu2025cashift}.

The current threat model assumes trusted collection, timestamps, parsing, and
labels. Because vocabulary and detector inputs derive from the scored stream, an
adversary influencing an unlabeled target catalog could alter vocabulary and IDF
before scoring without changing a label. An adversarial extension would examine
poisoning, injection, tampering, evasion, label corruption, and parser
compromise through controlled perturbations, catalog-provenance checks, and
\mbox{an explicit attacker model.}

\subsection{Documentation and reproducibility status}

The paper documents the protocols and the values they produce: split
construction, parser thresholds and snapshot points, temporal cutoffs,
high-level tokenization and vocabulary handling, the fixed hyperparameters,
visible catalog, and cost boundaries of Tables~\ref{tab:access},
\ref{tab:hyper}, and~\ref{tab:cost}, and the review-budget convention. The HDFS and BGL corpora are public and distributed with
their preprocessing provenance through Loghub~\cite{zhu2023loghub}. Release of a versioned experimental replication package containing split
indices, per-seed predictions, and table-regeneration scripts is planned
\mbox{as future work.}

\section{Conclusion}
\label{sec:conclusion}

Protocol choice changes the comparison among the six fixed configurations:
count/LR records the highest grouped mean AP on HDFS, semantic XGBoost the
highest chronological mean AP on BGL, and several configurations form a
near-ceiling cluster under random splitting. At a fixed BGL cutoff,
parser choice spans 0.124 in semantic XGBoost mean AP while preserving its lead
over count XGBoost;
the earliest rolling \mbox{period reverses that ordering.}

The cross-system ablation shows direction-dependent interactions between
catalog coverage and IDF weighting. Offline transductive access to unlabeled
target templates is associated with changes in reported AP and early review
queues, making pre-scoring information access
\mbox{an explicit experimental variable.}

Together, these findings identify split construction, parser state, catalog
visibility, review budget, and cost boundary as essential context for
interpreting results on HDFS and BGL. Reporting these properties alongside
scores supports protocol-aware comparison of detector accuracy and
\mbox{component costs.}

\section*{Acknowledgment}

Codex(OpenAI) was used for language editing, experiment running, and assistance in script development. The authors reviewed the manuscripts, verified the reported results and citations, and take responsibility for the final text.

\bibliographystyle{IEEEtran}
\begin{referencepage}
\bibliography{refs}
\end{referencepage}

\end{document}